\documentstyle[preprint,revtex]{aps}
\begin{document}

\draft
\centerline{Phys. Rev. E {\bf 64}, 1 November (2001)}
\vskip 20 mm
\begin{title}
Proteinlike behavior of a spin system near the transition\\
between ferromagnet and spin glass
\end{title}

\author{Chai-Yu Lin$^1$, Chin-Kun Hu$^{1}$, 
 and Ulrich H.E. Hansmann$^{2}$}

\begin{instit}
$^{1}$Institute of Physics, Academia Sinica, Nankang, Taipei 11529, Taiwan
\end{instit}

\begin{instit}
$^2$Dept.~of Physics, Michigan Technological University,
         Houghton, MI 49931-1295, USA
\end{instit}



\abstract{
A simple spin system is studied as an analog for proteins.
We investigate how the introduction of randomness and frustration
into the system effects the designability and stability of 
ground state configurations. We observe that the spin system exhibits 
protein-like behavior in the vicinity of the transition between 
ferromagnet and spin glass.
 Our results illuminate some guiding principles in protein evolution.}


\newpage

The folding of a protein into a specific three-dimensional (3D) 
biologically active structure is now often described 
by  
the funnel concept \cite{OLSW}. 
It is assumed that the energy landscape of a 
protein is rugged but with a sufficient overall slope 
towards the native structure \cite{DC}. 
Folding occurs by a multi-pathway kinetics and 
the particulars of the folding funnel determine the transitions
 between the different thermodynamic states \cite{DC,Bryngelson87}.
  While originally derived from  studies of minimalistic protein 
models, evidence for the validity of the funnel concept was
subsequently presented for real proteins \cite{HMO97b}.

A funnel-like energy landscape guarantees thermodynamic stability 
and kinetic accessibility for the biologically active structure of 
proteins. Both are necessary conditions for proteins to perform 
their biological functions. Hence, the funnel concept suggests 
that the optimal state of a protein is one of minimal frustration \cite{Go}. 
This is because a  smoother  energy landscape and a steeper slope 
leads to faster folding and greater stability.  However,  proteins 
are in general only marginally stable \cite{Privalov}, and both 
stability and speed of folding can often be increased in 
protein engineering \cite{Improve}. Hence, it appears that the sequence of
amino acids in a protein is in general not optimized for 
smoothness of its energy landscape. 
The question arises then on why is this the case and 
why are proteins only marginally stable. Or what factors constraint 
the amount of frustration (and  the ruggedness of the funnel 
landscape) in the evolution of proteins?

When studying the above questions  one encounters the problem that the 
amount of frustration  is difficult to control in protein models.  For 
this reason, we propose to use the frustrated 3D-Ising model on a 
simple  cubic lattice \cite {kirkpatrick77} with periodic boundary
conditions as an analogy of proteins, 
and to study the above questions for this much simpler system  in which 
the frustration  can be easily measured. Unlike in earlier 
work \cite{GHC,HSLC} we interpolate  continuously  between the ferromagnet 
and the spin glass by varying the density of antiferromagnetic bonds. 
Our choice of the system is motivated by the observation
that proteins are similar to spin glasses in that their energy 
landscape is characterized by a huge number of local minima 
separated by high energy barriers \cite{hu79}.  On the other hand, 
the global funnel-like topology of protein energy landscapes, leading 
to an  unique ground state, resembles more  a ferromagnet. Hence, it 
seems that proteins show behavior between that of a ferromagnet and a 
spin glass. However, the limitations of the analogy between the 
frustrated Ising model and proteins should be kept in mind. Spin systems 
do not fold, only the process by which the system finds its ground state 
can be regarded as analogous to folding. We  can study only how for
Ising models this process depends on the frustration and 
under which conditions there are similarities to proteins.

Our model is described by the Hamiltonian
\begin{equation}
 H = - \sum_{<lm>}^{3N} J_{lm} \sigma_l \sigma_m
\label{Ha}
\end{equation}
where the sum goes over all $3N$ ($N$ the number of spins) 
pairs $<lm>$ of nearest neighbor spins 
 $\sigma_i = \pm 1$. A certain number $M$ of randomly chosen 
 bond variables, $J_{lm}$, are set to $J_{lm}= -1$ while the 
remaining $3N-M$ bonds are assigned the value $J_{lm} = 1$. 
 The ratio $ R = {M}/{3N}$ 
is a measure for the randomness in our Ising system and leads to
the frustration in the systems which is as usual defined through
\begin{equation}
F = \frac{1}{3N} \sum_{\Box_i} \delta (F_{\Box_i}, -1) 
     \quad \quad {\rm with} \quad \quad 
     F_{\Box_i} =  J_{12}J_{23}J_{34}J_{14}~. 
\label{Fu}
\end{equation}
Here, $J_{12}$,$J_{23}$,$J_{34}$,$J_{14}$ are the four bond variables of 
the $i$-th
elementary plaquette $\Box_i$ of the lattice, and the sum goes over
 all $3N$
elementary plaquettes.

Our simulations are done on a $4 \times 4 \times 4 $  lattice 
which is small enough
that simulated annealing will  find the ground state. An even
smaller lattice size may have allowed
exhaustive enumeration, but would have introduced severe finite size effects. 
For a given value  $F$ of frustration, $2000$ realizations of bond 
variables $\{J_{lm}\}$ are generated in random. 
For each realization, $N_1$ simulated annealing runs are  used
to search for the global minimum. In each run
we cool down the system with step size $\Delta T$=0.1 from 
temperature $T$=3 to $T$=0.3 performing
40 Monte Carlo sweeps (one update for each spin) at  each temperature.
We define  as
ground state $C_g$ of one realization the configuration with
the lowest energy $E_g$ obtained in the  $N_1$ runs. 
To ensure reasonable statistics, we require
that this energy is found in at least $N_2$ simulated annealing runs. 
The total number $N_1 $ of runs is adjusted accordingly  
and the failure rate 
$N_F=(N_1-N_2)/{N_1}$
defines an index for the difficulty to find the global minimum.
In the next step, we check  the $N_2$ ground 
state configurations for rotational and translational symmetries, 
and identify in this way
the number $N_g$ of {\it distinct} ground state configurations
found for the given realization. For small values
of $R$ we set $N_2=10,000$. As the system approaches the
spin glass, $N_g$  increases rapidly. Therefore, if $N_g >1000$,
we repeat the simulation with $N_2=100,000$ to obtain 
more accurate values for $N_g$.

By altering the frustration
 $F$ we can tune our system between a ferromagnet ( $F=0$) 
and a spin glass ($<F>_{av}=0.5$) and investigate the relation 
between  $F$ in the system and  the occurrence 
of protein-like behavior. Since the  native state of a protein is 
unique and commonly assumed to be its ground state, we define 
a realization $\left\{J_{lm}\right\}$  
as  protein-like if it has a single ground state. The number of
protein-like realization $\left\{J_{lm}\right\}$ among
$2000$ realizations is denoted by $N_{SG}$. We display 
 the  frequency $f_{SG} = N_{SG}/2000$ of such
 realizations as a function of $F$  
in Fig.~\ref{fig1}, 
which shows that $f_{SG}$  decreases with growing $F$  
and is almost constant for $F \ge 0.44$. The inset of Fig.~\ref{fig1}
shows the same quantity as a function of $R$ and here flattening
occurs for $R \ge 0.23$.  
Hence, the probability to find protein-like realizations 
decreases as a function of $F$ (or $R$). However, the total number 
of realizations is  given by 
$  N_{Realizations} = {(3N)!}/[{(3N(1-R))! (3NR)! }]$,
i.e.  grows much faster with increasing $R$. It follows 
that the total number of protein-like realizations  
which can be designed (a randomly chosen realization has vanishing
small probability for a single ground state!)
is also  an
{\it increasing} function of $F$  since the bond randomness R and 
the average frustration over realizations $<F>_{av}$ are related through 
$<F(R)>_{av} = 4((1-R)^3R + (1-R)R^3)$ \cite {kirkpatrick77}.

We know  that  with growing $F$   the energy 
landscape  becomes more and more rugged.  The number of local minima
separated by high energy barriers will grow, and the probability will
increase that our simulated annealing runs get trapped in one of them 
and do not find the global minimum. This can be seen in
Fig.~\ref{fig2} 
where we display the average failure rate $<N_F>$ as a function
of $F$ for the case of all 2000 samples and for only
these realizations with single ground state $N_g=1$. 
In this plot we observe a steep increase  of 
 $<N_F>$  at $F_g= 0.44 \pm 0.02$ 
for the curve corresponding to the ``all samples'' case. 
Note that this value corresponds to $R_g=0.23 \pm 0.02$
which is consistent with that for the transition between
ferromagnetic and spin-glass order found in 
\cite{Hartmann}.  
The transition between the ferromagnet and the
spin glass can also be observed 
in the average number of ground states per realization
 $<N_g>$ as a function of $F$ which we  display in the 
inset of Fig.~\ref{fig2}. 
The location of the steep increase
in this quantity, $F_g=0.44 \pm 0.02$ (which corresponds to
 $R_g = 0.23 \pm 0.02$), 
is the same as for the failure rate and agrees with  
the point  in \cite{Hartmann}. 

The failure rate $N_F$ in Fig.~\ref{fig2} measures how often a simulation 
did {\it not} find the ground state and is therefore related to 
the ``folding time'',  i.e. the time which would be necessary  
to find the ground state in a 
simulation. The ``folding time'' itself is a measure for the 
kinetic accessibility of the ground states. 
For the frustrated Ising model we see from Fig.~\ref{fig2}
that the  failure rate (and consequently  the ``folding time'')
is small for small $F$
and differs little from the time needed for the ferromagnet $F=0$.
This changes once we reach values of  $F$ where the system 
behaves as a
spin glass. At that point the failure rate and the ``folding time''
 increases by orders of magnitude, and even for realizations
 $\left\{J_{lm}\right\}$ which have a single ground state, 
that state may no longer be kinetically accessible.
Such a situation is not  desirable for  real proteins,
which  have only limited time to fold and therefore 
must have kinetically  accessible native states.
Hence, we can not assume that realizations $\left\{J_{lm}\right\}$ with
$F \ge 0.44 \pm 0.02$ are protein-like even if they have a 
unique ground state. If the analogy between proteins and spin systems holds,
then we can expect  for  proteins also
 an interplay between the increasing entropy of sequences, which
lead to an unique ground state structures, and the requirement that
this state has to be kinetically accessible. On one hand the entropy
of sequences increases with frustration while on the other hand
the folding times become prohibitively large once the frustration
exceeds a certain value. In the 
Ising model the transition to this spin glass behavior is
pronounced and located at 
$F_g=0.44 \pm 0.02$ ($R_g = 0.23 \pm 0.02$). The above conjecture 
may explain why proteins are marginally stable: the entropy of 
marginally stable proteins is much higher than that of sequences 
optimized for thermodynamic stability and fast folding. However, a 
limiting minimal amount of thermodynamic stability
is necessary to guarantee function of the protein. 

The above conjecture implies that 
the ``optimal'' amount of frustration  in proteins is where the system
is ``almost'' at the point of becoming a spin glass. This is because in
such a case  the entropy of sequences which lead to a single {\it and} 
accessible ground state is maximal. However, 
%
a protein should also be stable  in the sense that a  mutation
will not 
lead to  an amino sequence with a {\it different} native structure
or no unique ground state at all. 
Hence, such protein structures are  
preferred which can be realized by a maximal number of different 
amino acid sequences \cite{tang}. In the language of our spin system
the above statement implies that these spin configurations are most 
protein-like 
which are single ground state for the largest number of 
realizations $\left\{ J_{lm} \right \}$. For this reason, we have 
further checked the $N_{SG}$ protein-like ground state configurations 
on translational and rotational symmetries. This procedure
leads  to a much smaller number $N_D$ of  distinct single ground state 
configurations. $N_D$ is displayed as a function of $F$ 
in the inset of Fig.~\ref{fig3}. $N_D$ is an increasing function 
over the whole ferromagnetic range and more or less constant in the
spin glass range.
Hence, with increasing value of  $F$ 
not only the total number of protein-like realizations grows but
also the variety of protein-like states.

>From the inset of Fig.~\ref{fig3} we would expect that
the situation in proteins would correspond
to small values of frustration $F$  in the  Ising model where
one  single ground state 
configuration dominates, which can be realized by many sets of bond 
variables $\left \{J_{lm}\right\}$. However, proteins have to change 
over the course of evolution. The requirement of evolutionary flexibility
suggests that larger values of randomness and frustration
should be preferred which increase 
the number of distinct ground state structures 
and  enhance the  chance that a mutation will lead 
from one structure to different one. Hence, we  expect for proteins
an interplay between the requirement that the native structure 
is stable under mutations, and the need for  structural changes
over the course of evolution.

In order to study this interplay we plot in Fig.~\ref{fig3} the 
ratio  $N_D/N_{SG}$. Note that this ratio corresponds to the inverse of
the (averaged) ``designability'' \cite{tang} and is a 
measure for the degeneracy of the various
protein-like
states (i.e. spin-configurations which are unique ground states 
for some  realizations $\left\{J_{lm}\right \}$)  of our spin system.
We see that this ratio has a step-like behavior at  
$F_p = 0.41 \pm 0.02$ (which corresponds to $R_p = 0.17 \pm 0.02$).
For smaller values of $F$  the $N_D$ types of 
ground state 
configurations are realized by many sets $\left\{J_{lm}\right\}$, 
while for larger values of $F$ 
 each spin configuration is realized  by only one
realization $\left\{J_{lm}\right\}$. Hence, we conclude that 
in our spin system the ``optimal'' frustration is at $F_p$ 
where  both  a variety of different  protein-like configurations
can be realized, but at the same time these structures 
can be designed by more than one set of $\left\{J_{lm}\right\}$, 
and therefore are stable under mutations \cite{footnote}. 
Note that this point is close to,
but smaller than, the glass transition point ($F_g=0.44 \pm 0.02$).
Our value of $F_p$ also corresponds  to the point where in
Fig.~\ref{fig1}  failure rate of  realizations with single ground 
state diverges from the corresponding plot for {\it all} 
realizations: $F = 0.41 \pm 0.03$.

The above results suggest  that in protein-like
systems randomness and frustration is necessary to increase 
the designability of proteins. In our spin system, the absolute
number of realizations with a single ground state will increase with
frustration. 
On the other hand,
once the frustration  exceed a certain value, the system
becomes a spin glass. The resulting rugged energy landscape implies now
that the single ground state, if existing, is no longer kinetically
accessible. 
This would be biologically not desirable, and the frustration in proteins
has to be below this critical value.
In a similar fashion, the evolutionarily favorable increase in diversity
of protein-like states  with  frustration is 
counteracted by the growing probability that a given configuration 
becomes unstable under mutations.  If the frustration exceed a 
certain value, any mutation would lead 
to a different structure which is again biologically not desirable.
We conjecture that  proteins are not minimal frustrated but
that in protein-like systems the competition between these 
factors leads to a maximal value of $F$  where  the number of
different kinetically accessible structures, which can be realized as
single ground states by {\it many} sequences, is largest. 
For our spin system this point is 
$F_p = 0.41 \pm 0.02$ which is
close to, but below the point $F_g=0.44 \pm 0.02$ 
where the system starts to behave as a spin glass.

In order to demonstrate how the interplay of the above outlined factors 
may lead to an optimal value of $F$, we have made up the following game.
Our starting point is the ferromagnet, i.e. $J_{lm} = 1$.  The game 
consists of a series of Monte Carlo steps which simulate ``evolution''.
At each Monte Carlo step our system has two offspring
before it dies. One of the offspring is a copy of the parent, the other
carries a mutation. We simulate mutations by 
chosing at random  one bond variable $J_{lm}$ and switching its sign.
Only one of the offspring is allowed to survive, and the survival rate
of the ``mutant'' is given by  $P(F_N)/(P(F_N)+P(F_0))$.  Here, $F_N$ and
$F_0$ are the frustration of the ``mutant'' and the ``unchanged system'',
respectively, with
$  P(F) = f_{SG}(F) (1-<N_F(F)>) (1 - {N_D(F)}/{N_{SG}(F)})$,
where $f_{SG}(F)$, $<N_F(F)>$,  and $N_D/N_{SG}$ are taken 
from our previous simulations and $<N_F(F)>$
corresponds to the curve $<N_F(F)>$ with $N_g=1$ in Fig.~\ref{fig1}.   
With these rules our system performs a random walk in $F$
shown in Fig.~\ref{fig4}. The average value of $F$ throughout this
random walk gives  $F=0.42 \pm 0.03$
which is consistent with $F_p=0.41 \pm 0.02$
and supports  our assumption that the evolution
of protein-like systems leads to a optimal point of $F$ 
in the system. 

In summary, we have studied the simple frustrated Ising model as 
an analog for proteins. Investigating this system as a function of 
frustration, we found that the spin system exhibits protein-like 
behavior at or slightly below the point at which a system changes 
from an ordered (ferromagnet) to a random system (spin glass). 
Whether this observation (which questions the common belief that
proteins are minimal frustrated systems) holds for realistic protein 
models remains to be investigated.  As a next step in this direction 
we have started simulations of a bond-diluted and site-diluted 
frustrated Ising model.
In such a  model, it may be
possible to generate more realistic protein-like structures with 
backbone and side chains.

 We thank Jonathan Dushoff and C. Tang for a critical reading of the paper.
U.H. acknowledges support by a research grant (CHE-9981874) of
the National Science Foundation. This work is also supported by the
National Science Council of the Republic of China (Taiwan) under Grant No.
89-2112-M001-084.


\figure{The frequency $f_{SG}=N_{SG}/N_T$ of realizations 
 with single ground state 
  as a function of $F$ and (inset) $R$.
\label{fig1}}

\figure{The  average failure rate $<N_F>$ as a function of $F$. In the inset
       we display  the average number $<N_g>$ of ground states as a function 
       of $F$.
\label{fig2}}

\figure{The ratio $N_D/N_{SG}$  as a function of $F$. In the inset we show 
   the number $N_D$ of {\it truly different} single ground state 
         configurations, as a function of $F$.
\label{fig3}}

\figure{Time series of the bond randomness $F$  from a dynamic simulation
    described in the text. 
\label{fig4}}

\end{document}